\documentclass[aps,superscriptaddress,showpacs]{revtex4}
\usepackage{graphicx}

\renewcommand{\v}[1]{{\vec{ #1}}}

\newcommand{\s}{{\sigma}}

\def\be{\begin{eqnarray}}
\def\ee{\end{eqnarray}}
\newcommand{\nn}{\nonumber\\}

\newcommand{\Eq}[1]{Eq.~(\ref{#1})}
\newcommand{\Fig}[1]{Fig.~(\ref{#1})}
\newcommand{\e}{\epsilon}

\newcommand{\ra}{\rightarrow}

\begin{document}

\title{What Makes the $T_c$ of FeSe/SrTiO$_3$ so High ? }
\author{Dung-Hai Lee}
\affiliation{Department of Physics, University of California at Berkeley, Berkeley, CA
94720, USA}
\affiliation{Material Science Division, Lawrence Berkeley National Laboratory, Berkeley,
CA 94720, USA}
\date{\today}
\begin{abstract}
This note reviews some of the recent progresses in the study of high temperature superconductivity in the interface between a single unit cell FeSe and SrTiO$_3$. It offers the author's personal view of why $T_c$ is high and how to further increase it. 
\end{abstract}
\maketitle
Raising the superconducting transition temperature ($T_c$) to a point where applications are practical is one of the most important challenges in science. In the history of high $T_c$ superconductivity there are two landmark events -- the discovery of copper-oxide superconductor in 1986\cite{BM}, and the discovery of iron-based superconductor (FeSC) in 2006\cite{Hosono}.\\

For the copper-oxide superconductors the higest $T_c$, under ambient pressure, is achieved in  HgBa$_2$Ca$_2$Cu$_3$O$_8$ ($T_c=134K$). For the FeSCs $T_c$ has been raised to $~55K$ in ${\rm SmO}_{1-x}{\rm F}_x{\rm FeAs}$ in 2008\cite{Ren}. This sets the record for FeSCs until quite recently.  In 2012 an anomalously large superconducting-like energy gap was seen by scan tunneling microscopy (STM) in an one unit cell (1UC) thick FeSe film grown on the TiO$_2$ terminated (001) surface of SrTiO$_3$ (STO)\cite{Xue}. This report stirred up lots of interests because the observed energy gap invites the optimism that the transport $T_c$ is very high.\\

Subsequent angle-resolved photoemission spectroscopy (ARPES) studies have shown that this superconducting-like energy gap opens at temperatures range from 55K to 75K\cite{XJZ1,XJZ2,DF1,DF2,JJ} (the value depends on the precise substrate, e.g.,whether it is STO or BaTiO$_3$, and the growth condition).
However the anticipated high transport $T_c$ has not been established until, possibly, two recent experiments. 
The first is an in-situ four-probe transport measurement which shows the onset of superconducting $I-V$ characteristics at temperature higher than 100K\cite{Jia}. The second is a recent mutual inductance measurement showing a $\sim$ 60K diamagnetic transition in a sample where the ARPES gap opens around 65K\cite{yayu}.Of course both experiments need further confirmation.\\

From the ARPES point of view the observed energy gap behaves as what one expects for the superconducting gap. (1) The node-free energy gap is adhered to the Fermi surface and is particle-hole symmetric\cite{XJZ2}. This means along momentum cuts going through the Fermi surface the energy gap minimum always occurs on the Fermi surface.  (2) The energy gap evolves with temperature as an order parameter does in a second order phase transition\cite{XJZ1,XJZ2,DF1,DF2,JJ}. The particle-hole symmetry makes it unlikely that the gap is due to particle-hole (rather than particle-particle) condensation (e.g., certain kind of density-wave order). Because typically in such kind of order it requires the breaking of translation symmetry to open energy gap. And to produce a particle-hole symmetric nodeless gap the Fermi surface needs to be completely nested by the ordering wavevector(s). Such nesting is not observed in the measured Fermi surface\cite{XJZ1,XJZ2,DF1,DF2,JJ}. Although this 
issue deserves further scrutiny, in the rest of the paper I shall {\it assume} the energy gap seen by ARPES is  due to Cooper pairing, and when discussing ARPES results we use ``$T_c$'' to denote the gap opening temperature.\\
\begin{center}
\begin{figure}
\includegraphics[scale=0.3, angle=0]{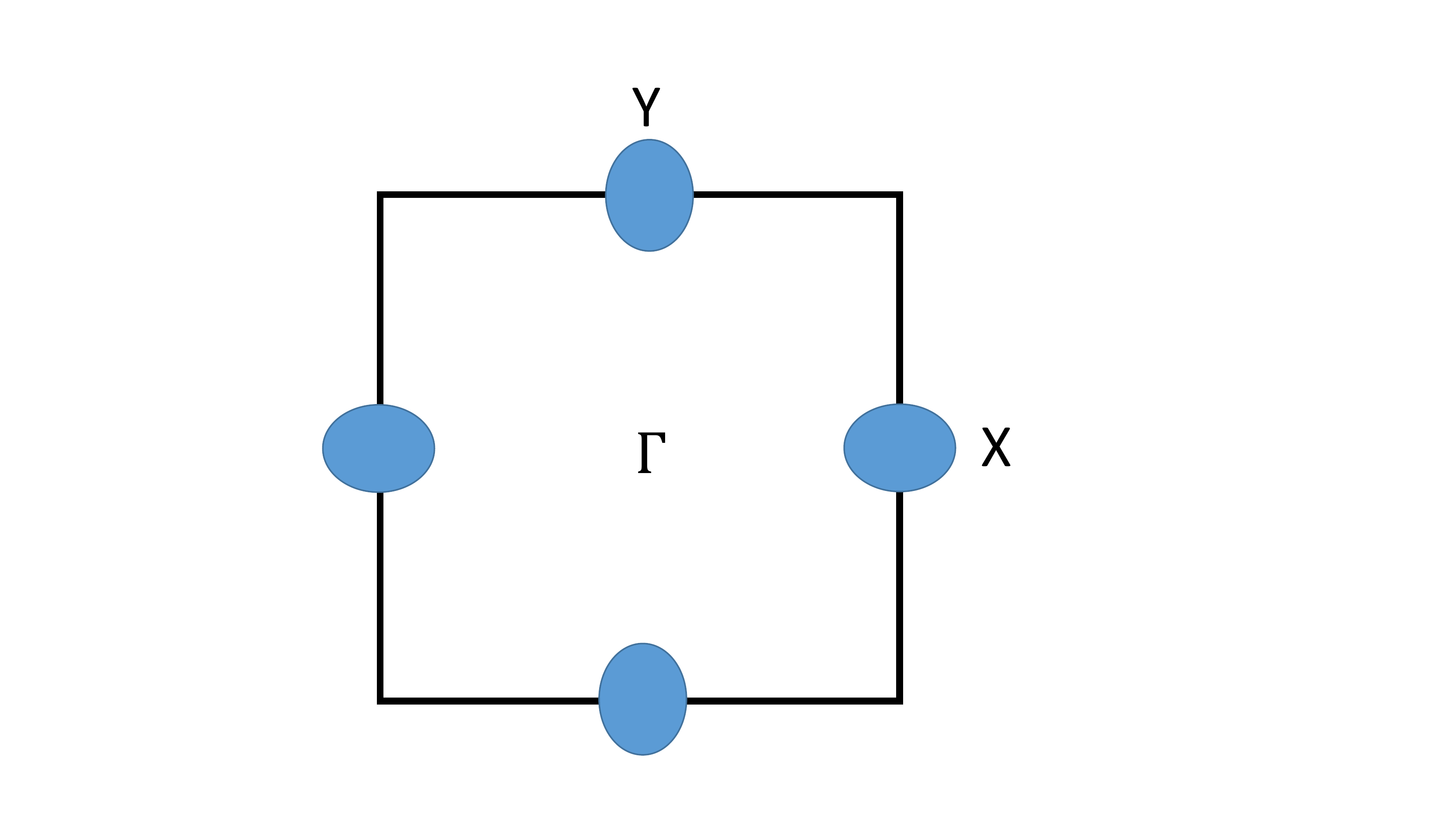}
\caption{A schematic illustration of the Fermi surface of 1UC FeSe/SrTiO$_3$. Here the Brillouin zone corresponds to a real space unit cell enclosing one Fe atom.}
\label{FS}
\end{figure}
\end{center}

\begin{center}
\begin{figure}
\includegraphics[scale=0.2, angle=0]{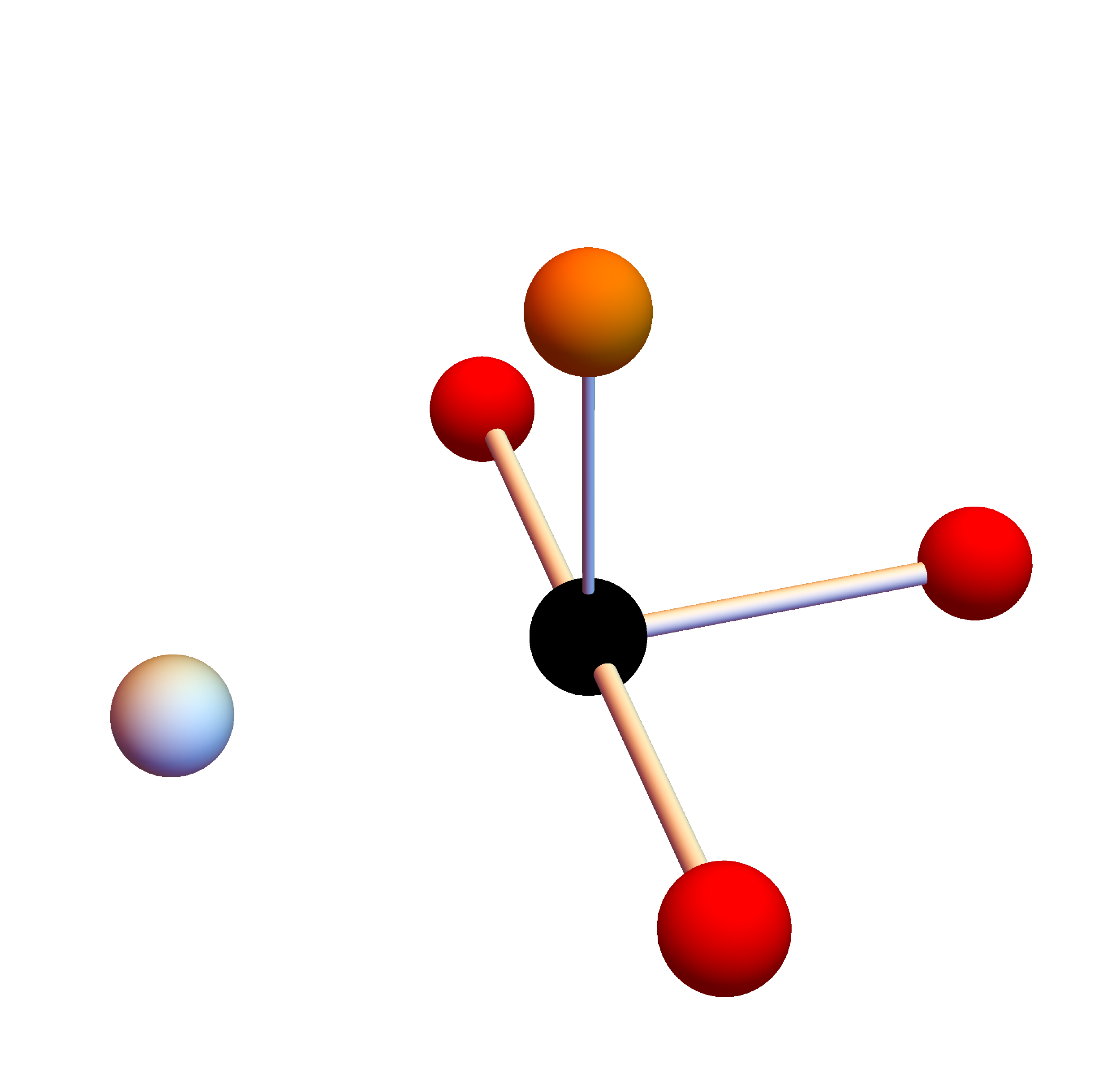}
\caption{A schematic illustration of a Ti(black) neighboring three in-plane oxygens (red) and an oxygen vacancy (gray). It transfers some charge hence establishes a weak bond with the Se atom (orange) above it. }
\label{bond}
\end{figure}
\end{center}

The Fermi surface of 1 UC FeSe film on STO shows only the electron pockets. This is schematically shown in the ``unfolded'' Brillouin zone (i.e. the Brillouin zone corresponding to a real space unit cell enclosing one iron atom) in \Fig{FS}. 
In particular, the hole pockets at the Brillouin zone center, which are observed in other FeSCs, are absent. Because the presence of these hole pockets are demanded by charge neutrality (since the  stoichiometric FeSe has 2 valence electrons per unit cell), the lack of them implies the FeSe film is {\it electron doped}.  It is widely believed (but not yet proven) that the doping originates from the oxygen vacancies present on the STO side of the interface.
Microscopically one can imagine a oxygen-vacancy-bordered Ti atom donating some (averaged number of) electrons to the Se above it (see \Fig{bond}). \\

The Fermi surface of 1 UC FeSe/STO is very similar to that of A$_{x}$Fe$_{2-y}$Se$_2$\cite{DF3,XJZ3} (where A=K,Cs,Rb,Tl..etc, and $T_c\sim 30K$) as well as that of (Li$_{1-x}$Fe$_x$)OHFe$_{1-y}$Se ($T_c=41 K$)\cite{xh} determined in a recent ARPES experiment\cite{zhao}. In addition, it is also very similar to the Fermi surface of the potassium (K) doped 3 UC FeSe/STO ($T_{c,max}\sim 48K$)\cite{miyata}. The fact that strong Cooper pairing exists in systems without the hole pockets definitely rules out theories where the approximate nesting of the electron and hole pockets is important for 
high temperature superconductivity.\\

When comparing the $T_c$ of (a) A$_{x}$Fe$_{2-y}$Se$_2$ ($\sim$ 30K) (b) (Li$_{1-x}$Fe$_x$)OHFe$_{1-y}$Se (41K)(c) the K-doped 3 UC FeSe/STO (48K) and (d) the 1UC FeSe/STO (55-75K) with that of bulk FeSe ($\sim 8$K) it is important to keep the following facts in mind. The first is (a)-(d) all show similar electron-pocket-only Fermi surface in contrast to bulk FeSe which has both electron and hole pockets. This fact suggests {\it electron doping enhances $T_c$}. Second, despite both being FeSe thin films on STO and  have  very similar electron pockets, the maximum gap of 1UC FeSe/STO (14-18 meV) is significantly larger than that of the K-doped 3UC FeSe/STO ($\sim$ 9 meV)\cite{miyata}. This suggests that electron doping alone can not account for the full enhancement of Cooper pairing strength in 1 UC FeSe/STO. Furthermore it is important to note that for K doped 3 UC FeSe/STO the doping is achieved through the charge transfer from K, while for the 1 UC FeSe/STO the doping comes from the STO substrate. In later discussions we shall argue that the FeSe $\leftrightarrow$ STO charge transfer is critically important for the extra $T_c$ enhancement. \\

In addition to what have already been discussed, the following facts are also useful hints for the $T_c$ enhancement mechanism of 1 UC FeSe/STO. (1) STO is a special substrate, i.e., just thinning FeSe to 1 UC thick is not sufficient to give rise to high $T_c$. Indeed when thin FeSe films were grown on, e.g.,  graphene/SiC(0001), the  maximum $T_c$ is that of bulk FeSe, and is reached in thick films\cite{SIC}.  (2) The fact that the a-b lattice constant of (Li$_{1-x}$Fe$_x$)OHFe$_{1-y}$Se is nearly the same as bulk FeSe suggests tensile strain is not the reason for $T_c$ being enhanced from 8K to 41K. Moreover by carefully changing the detailed properties of the interface Ref.\cite{DF2} concluded that tensile strain only has a small effect of $T_c$ . (3) The effective mass of the Fermi-crossing electron band in 1UC FeSe/STO is about the same as that of (Li$_{1-x}$Fe$_x$)OHFe$_{1-y}$Se but both are considerably smaller than that of (Tl,Rb)$_x$Fe$_{2-y}$Se$_2$\cite{zhao}. However the $T_c$ of the former two materials are higher than the latter. This speaks against the high $T_c$ in 1 UC FeSe/STO is due to a  stronger electron-electron correlation.\\

\begin{center}
\begin{figure}
\includegraphics[scale=.45]{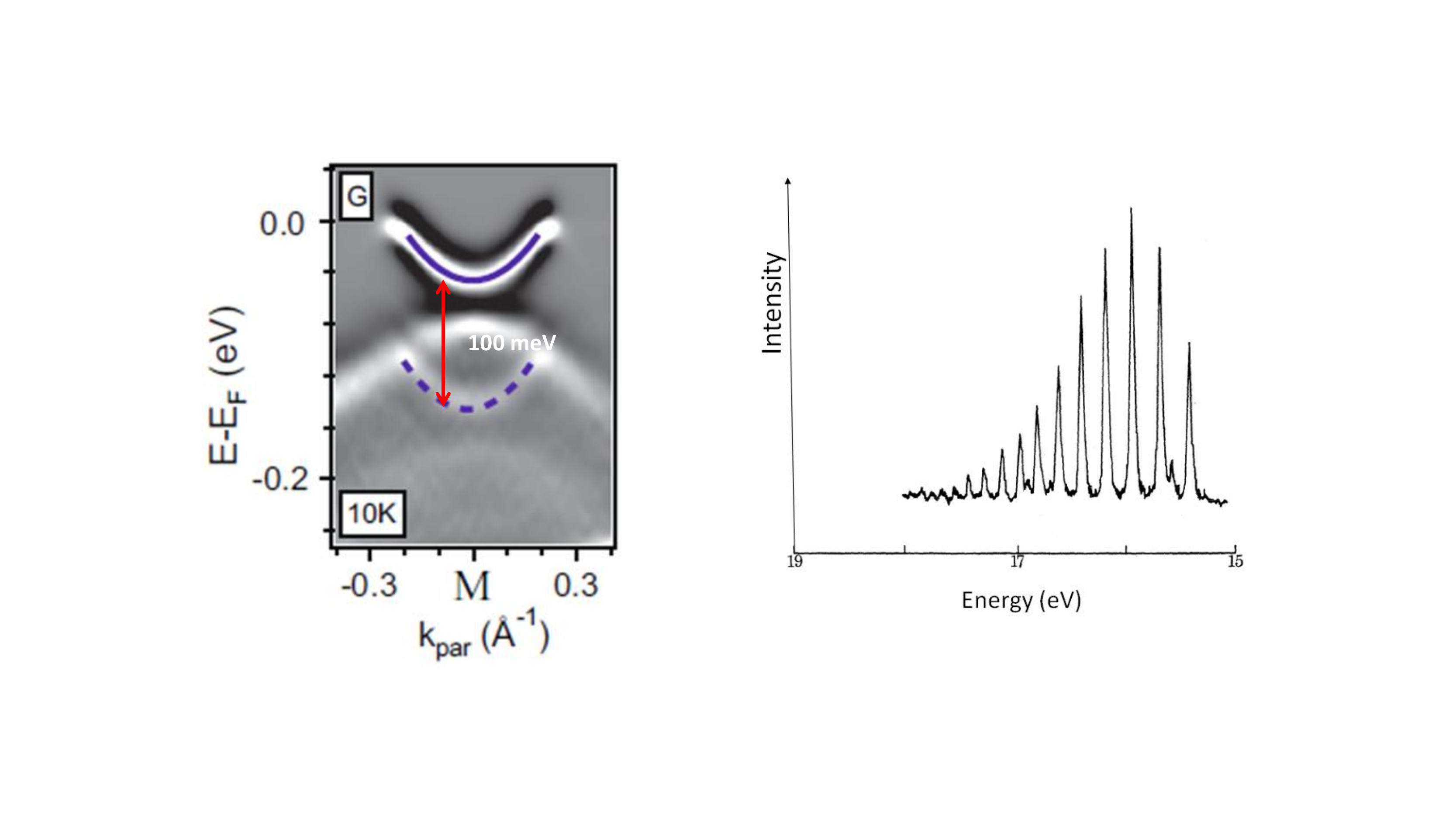}
\caption{Left panel: reproduction of data from Ref.\cite{JJ}. What's shown is the second energy derivative of the ARPES intensity as a function of momentum around the M point of the Brillouin zone. (The M point is the equivalence of the $X$ and $Y$ points in \Fig{FS}. ) Right panel: A reproduction of the angle-integrated photoemission data of H$_2$ molecular gas from Ref.\cite{H2}.}
\label{image}
\end{figure}
\end{center}

The foremost telling clue for the origin of the $T_c$ enhancement in 1 UC FeSe/STO comes from the ARPES results 
reported in  Ref.\cite{JJ}. Similar to earlier findings a maximum gap $\sim$ 14 meV was observed and 
the gap closes at $T_c\sim 58$K. 
However the most striking finding is the observation of ``replica bands''. In the left panel of \Fig{image} we reproduce the data from Ref.\cite{JJ}. A replica of the parabolic electron-like band is observed $\sim$100 meV away in the higher biding energy direction. Note that despite it is situated far below the Fermi energy the replica band stops where the main band crosses the Fermi level.  In addition, a faint replica of the non-Fermi-level-crossing hole-like band can also been seen. Again, the energy separation between the main band and the replica band is approximately 100 meV. Moreover as the main band disperses toward the Fermi energy it ``bends back''. This is a characteristic phenomenon associated with the opening of an energy gap on the Fermi surface. The same back bending is weaker but observable for the replica band ! Lastly high statistics ARPES data\cite{JJ} (not shown here) even shows the second replica band 200 meV away from the main band.\\

To understand the origin of the replica band, in the right panel of \Fig{image} we reproduce the angle-integrated photoemission data of the H$_2$ molecular gas reported in Ref.\cite{H2}. Interestingly, in addition to the main photoemission peak there are a series of replica peaks. A closer examination revealed that the energy separation between the adjacent peaks is the energy of the bond stretching vibration quantum. The consensus attributes this phenomenon to the ``vibron shake off'', i.e., each replica peak corresponds to photo emitting an electron plus exciting an integer number of vibration quanta.

Comparing the H$_2$ replicas with those observed in FeSe/STO motivated us to think the replica bands are also due to shaking-off of certain $\sim 100$ meV bosonic excitation. Interestingly pure STO has a very flat optical phonon band centered around 100 meV\cite{phonon1,phonon2,phonon3}. This phonon band is  primarily oxygen-vibration in character and is separated from other phonon bands by a substantial energy gap.  Naturally this motivates us to think the boson in question are these high energy STO phonons. The facts that the replica of the Fermi level crossing band terminates at the momentum where the main band crosses the Fermi energy, and in the superconducting state it bends back at the same momentum where the main band bends back, rule out other explanations. These explanations include the replica bands being the quantum well state,  or they are the main band rigidly shifted downwards in energy because the potential energy in certain poorly connected part of the sample is different from other regions.
Finally the strongest support of our phonon shake-off explanation is the recent observation of $\sim$ 100 meV replicas of the surface bands of pure STO\cite{wang}.\\

In addition to the replica bands there are several other  findings of Ref.\cite{JJ} that are significant. (A) The band structure of films with thickness  $\ge 2$ UC resembles that of bulk FeSe. In particular it shows both hole and electron pockets indicating these films are much less electron-doped. (The fact that only the 1UC film has significant charge transfer with STO is not understood at present time.)
(B) The replica bands are only observed for 1 UC FeSe film -- films with thickness $\ge 2$ UC do not show replica bands. These multi UC films also do not show any appreciable superconducting gap\cite{JJ}.  (C) The multi UC films show the splitting between the xz and yz bands near the M point which acts to reconstruct the electron pockets. Such band splitting is  characteristic of the 90$^o$ rotation symmetry breaking (i.e. nematicity) found in other iron pnictides\cite{Yi} and  bulk FeSe\cite{Coldea,donghui,Wei}. However the lack of band folding associated with the stripe antiferromagnetic (AF) order suggests, like bulk FeSe\cite{bulk1,bulk2}, these films are nematic but not stripe AF ordered. (D) The 1 UC film does not show the xz/yz band splitting or any sign of strong nematic fluctuation. The last statement is drawn from the fact that STM quasiparticle interference studies of 1UC FeSe/STO do not show any sign of $C_4$ symmetry breaking\cite{DL5}. We interpret this as indicating the weak 
response of the electronic structure to the inevitable anisotropic local environment around impurities -- hence the small nematic susceptibility/fluctuation\cite{yayu2}. \\

Fact (A) suggests the charge transfer from STO is only significant in the 1 UC film, and fact (B) suggests such charge transfer is important for the presence the replica band or strong electron-phonon coupling (see later discussions for an explanation). This explains why we do not expect strong coupling to the STO phonon in K-doped 3 UC film -- the doping is not due to the charge transfer from STO.   These arguments, plus  the fact that despite similar doping level the $T_c$ of the 1 UC film is significantly higher than that of the K-doped 3 UC film, suggests a connection between the strong coupling between the FeSe electron and STO phonon and the extra $T_c$ enhancement in the 1 UC film. Lastly facts (C,D) imply, unlike many other pnictides\cite{fk}, the highest $T_c$ sample has the weakest nematic susceptibility.
\\

\begin{center}
\begin{figure}
\includegraphics[scale=.4]{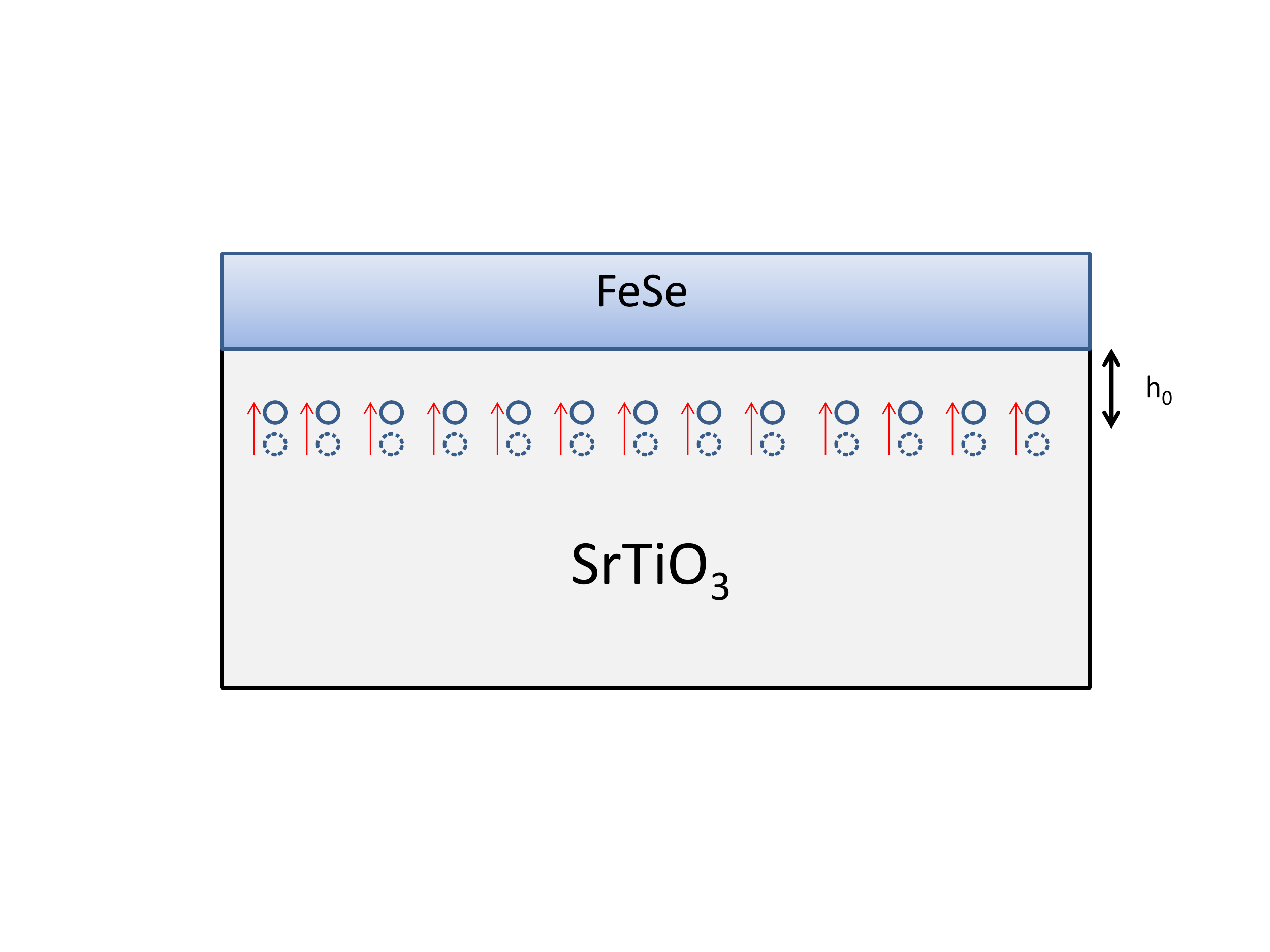}
\caption{The vibration of the oxygen anions in the interface dipoles exert a time-dependent dipole potential on the electrons in FeSe causing the electron-phonon coupling responsible for the replica bands. Here $h_0$ is the set back distance between the dipole layer and the Fe plane in the FeSe layer.}
\label{interfdipole}
\end{figure}
\end{center}

Since STO is almost a ferroelectric, it is natural to think the electric field set up by the charge transfer at the interface can induce a layer of ordered dipoles. Note that the free carriers in STO, due either to oxygen vacancies or the Nb doping (in all the ARPES studies the STO is doped with Nb), will screen the electric field away from the interface. Hence we expect the induced dipole layer to   localize near the interface. These dipoles are due to the relative displacements of the Ti cations and the oxygen anions (as in bulk STO). Hence the vibration of the oxygen anions will lead to a modulating dipole potential in the FeSe layer. A caricature of the situation is shown in \Fig{interfdipole}.

Let $\delta P_z$ be the change in the dipole moment due to the displacement of the oxygen anions (in direction perpendicular to the interface)
\be
\delta P_z(x,y,-h_0)=q_{eff}\delta h(x,y,-h_0).\ee Here $x,y$ are the coordinates in planes parallel to the interface, and the origin of the $z$ coordinate is chosen at the Fe plane in FeSe. Relative to the Fe plane the dipole layer sits at  $z=-h_0$. The induced change of the dipole potential in the Fe plane due to a ``frozen'' oxygen displacement is given by\cite{JJ} 
\be
\Phi(x,y,0)={\e_\parallel q_{eff} h_0\over \e_\perp^{3/2}}\int dx' dy'{\delta h(x',y',-h_0)\over \left({\e_\parallel\over\e_\perp} h_0^2+(x-x')^2+(y-y')^2\right)^{3/2}}.\ee
Fourier transform the above result with respect to $x,y$ we obtain
\be
\Phi(\v q_\parallel,0)=\sqrt{{\e_\parallel}\over\e_\perp}{2\pi q_{eff}\over \sqrt{\e_\perp}} e^{-|\v q_\parallel| h_0\sqrt{\e_\parallel/\e_\perp}}~\delta h(\v q_\parallel,-h_0).
\label{eph}
\ee
Here $\v q_\parallel$ is the wavevector parallel to the interface, and $\e_\parallel,\e_\perp$ are the dielectric constants parallel and perpendicular to the interface, respectively. They contain the contribution from both
STO and the FeSe film. Because the FeSe electrons are confined to move parallel to the
interface, they only contribute to $\e_\parallel$. 
Aside from the screening from the FeSe carriers we expect STO (which has cubic crystal symmetry in the bulk) to contribute about equally to $\e_{\parallel}$ and $\e_\perp$. This leads 
us to expect the total dielectric constant $\e_\parallel$ to be significantly bigger than $\e_\perp$. \\

\begin{center}
\begin{figure}
\includegraphics[scale=.25]{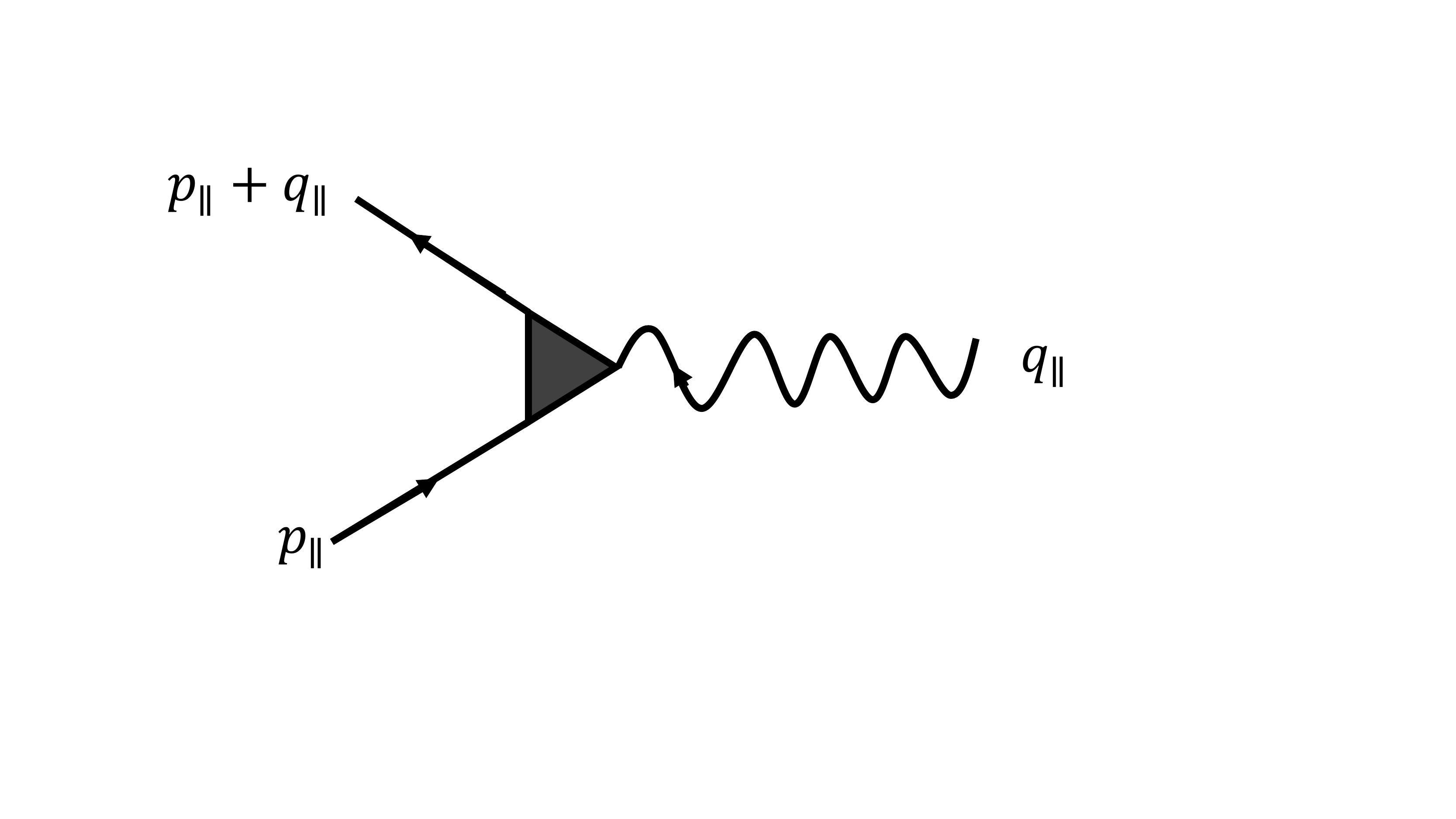}
\caption{The electron phonon interaction matrix element. The fact that the replica bands  follow the main band dispersion so closely suggest the electron-phonon matrix element strongly peaks at $\v q_\parallel=0$.   }
\label{ephme}
\end{figure}
\end{center}

\Eq{eph} leads to the following electron-phonon coupling term in the Hamiltonian
\be
&&\sum_{\v q_\parallel,\v k_\parallel,\s} \Phi(\v q_\parallel,0) \psi^+_{\v k_\parallel+\v q_\parallel,\s}\psi_{\v k_\parallel,\s},\nn
\ee
where $\psi$ annihilates electrons in FeSe.
Thus the electron-phonon coupling constant (see \Fig{ephme}) depends on $\v q_\parallel$ via \be &&\Gamma(\v p_\parallel,\v q_\parallel)=\sqrt{{\e_\parallel}\over\e_\perp}{2\pi q_{eff}\over \sqrt{\e_\perp}} e^{-|\v q_\parallel|/q_0}\nn&& q_0^{-1}=h_0\sqrt{\e_\parallel/\e_\perp}.\label{epc}\ee 

The fact that $\e_\parallel >> \e_\perp$  results in two important
effects: (1) the strength of the electron-phonon coupling is enhanced by
$\sqrt{\e_\parallel/\e_\perp}$;
(2) $q_0$ is reduced by a factor
$1/\sqrt{\e_\parallel/\e_\perp}$ resulting in the e-ph coupling function more sharply
peaked at $\v q_\parallel = 0$. \\

The forward-focusing electron-phonon interaction in \Eq{epc} explains why the replica band follows the dispersion of the main band so closely. Indeed, had the e-ph coupling  allowed a wide range of momentum transfer, the image of the main band would consist of the superposition of many replica bands each displaced from the main band by a different momentum. If so the dispersion of the replica band would have been blurred out.\\

Through the electron-phonon interaction discussed above the high energy phonon generates an effective electron-electron attraction. Because the energy of the involved phonon  ($\sim 100$ meV) is considerably larger than the electron bandwidth ($\sim 65$ meV)  we can simply ``integrate out'' the phonon to obtain an instantaneous effective attraction\cite{JJ}
\be
&&U_{eff}=
-\sum_{\v k,\v p,\v q}\sum_{\s,\s'}g(\v q)\psi^\dagger_{\v k+\v q,\s}\psi_{\v k,\s}\psi^\dagger_{\v p-\v q,\s'}\psi_{\v p,\s'}~~{\rm where}\nn &&g(\v q)= {v_{eff}\over 2\pi q_0^2}e^{-|\v q|/q_0}.\ee
(Here we have dropped the subscript $\parallel$.)
Such effective attraction can enhance the Cooper pairing arising from pure electronic mechanism under the proviso that ``for most'' $\v k$ points the gap function  resulted from the electronic  mechanism does not change sign as $\v k\ra\v k+\v q$ (here $\v q$ lies in the range where the electron-phonon coupling constant in \Eq{epc} is appreciable). Quantitatively when $U_{eff}$ is weak whether it  enhances or suppresses the pairing due to the electronic mechanism is determined by the following parameter\cite{JJ}
\be
\lambda = {\int_{FS} d\v k\int_{FS} d\v p~ \Delta^*(\v k)g(\v k-\v p)\Delta(\v p)\over \int_{FS} d\v k ~|\Delta(\v k)|^2}.\ee Here the momentum integrals are performed along the Fermi surface,and $\Delta(\v k)$ is the gap function caused by the electronic pairing.  $U_{eff}$ enhances (suppresses) pairing if $\lambda>0$ ($\lambda<0$). \\

\begin{center}
\begin{figure}
\includegraphics[scale=.5]{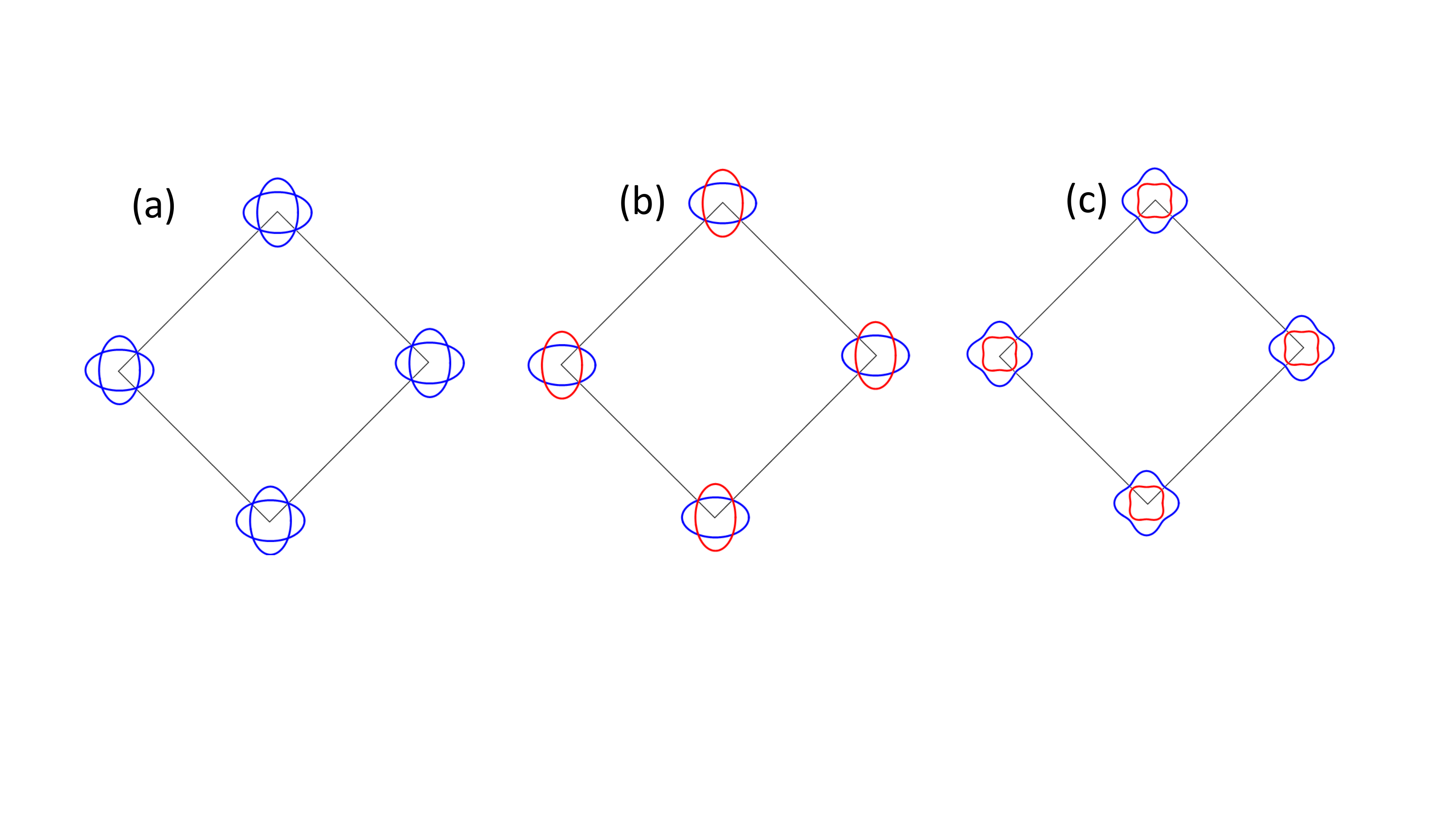}
\caption{Schematic representations of three scenarios for the gap function in FeSe/STO. Here the color designates the sign of the gap function (blue: + and red:-). It is interesting that the Fermi surface splitting at the crossing momenta (as depicted in panel (c)) is not observed in the ARPES result of Ref.\cite{yan}.}
\label{gapform}
\end{figure}
\end{center}

For 1UC FeSe/STO several scenarios for $\Delta(\v k)$ have been discussed in the literature. They are shown in \Fig{gapform}, where colors denote the sign (say, blue: positive, red: negative). In the case of panel (a) $\lambda$ is clearly positive and the electron-phonon interaction will enhance $T_c$. However for panel (b) and (c) the situation is not so clear. First of all, either the breaking of the glide plane symmetry (i.e., translation by a Fe-Fe distance followed by a reflection about the Fe plane) by the substrate or spin orbit interaction will split the two Fermi surfaces where they cross. This hybridization effect, even very weak, will induce gap nodes at the crossing momenta for scenario (b). Such gap nodes are not observed in recent high resolution ARPES measurement of the superconducting gap as a function of momentum\cite{yan}. Panel (c) where the gap function changes sign between the 
two hybridization-split Fermi surfaces is unfavored by recent STM study examining the change of quasiparticle interference as a function of applied magnetic field as well as the impurity induced in-gap states for scalar and magnetic impurities. The result suggests there is no sign change in the gap function\cite{DL5}. It is also unfavored by the best gap function that fits of the measured energy gap in ARPES\cite{yan}. Therefore we believe the $\Delta(\v k)$ of 1 UC FeSe/STO is qualitatively represented by panel (a) of \Fig{gapform}.        

\begin{center}
\begin{figure}
\includegraphics[scale=.4,angle =0]{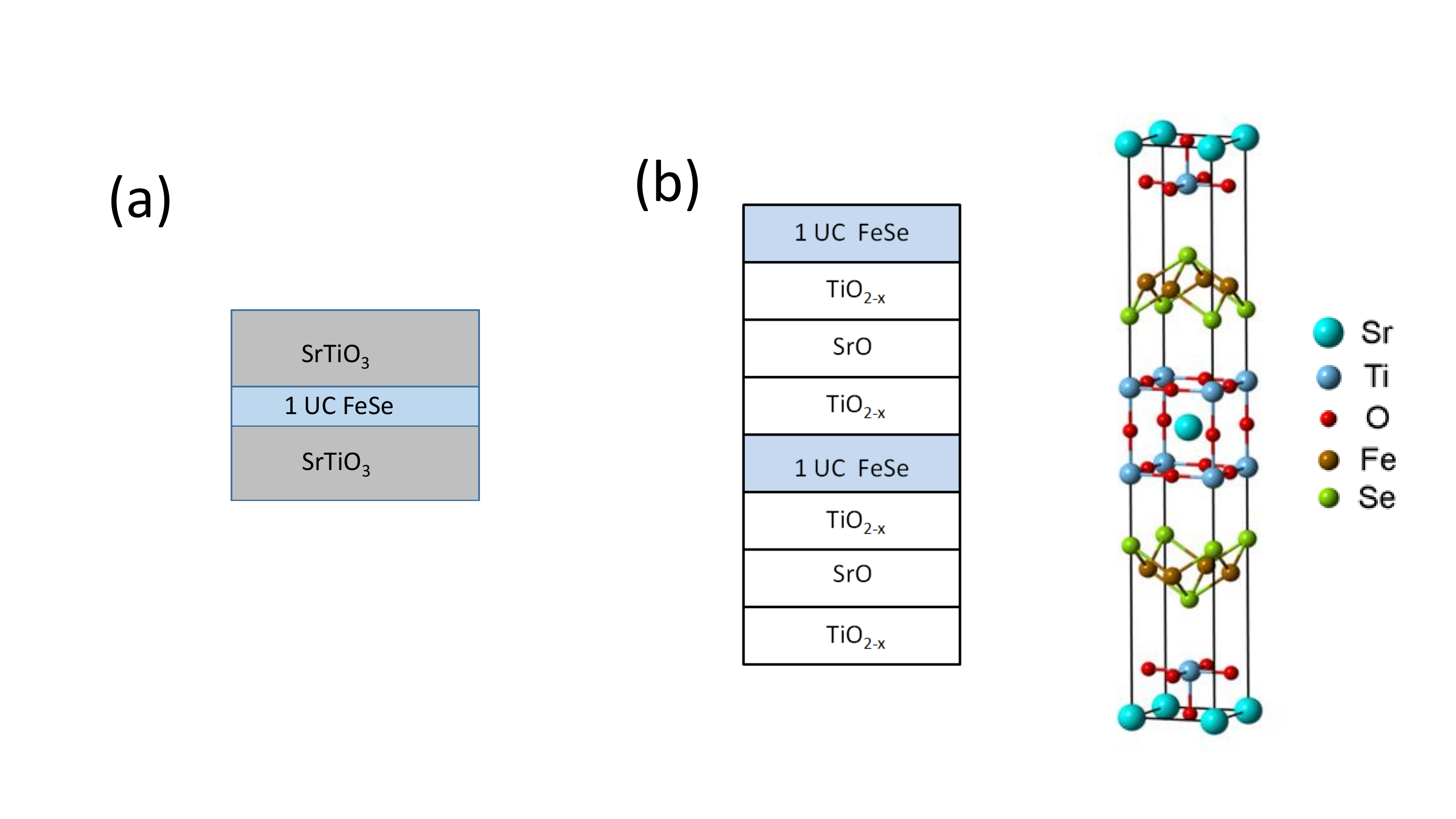}
\caption{The proposed high $T_c$ structures. In panel(a) FeSe makes interfaces on both sides with STO. Growing this structure by, say, molecular beam epitaxy is very challenging because the required growth temperature of STO can nearly melt FeSe. (b) A bulk crystal in which each FeSe layer is sandwiched between two TiO$_2$ layers. In order to ensure charge transfer it is important to maintain the existence of oxygen vacancies in the TiO$_2$ planes. }
\label{sand}
\end{figure}
\end{center}

To summarize what has been discussed so far, the following two mechanisms cooperatively contribute to the high $T_c$ of the 1 UC FeSe/STO:
\begin{enumerate}
\item A pure electronic pairing  mechanism which alSO operates in  A$_{x}$Fe$_{2-y}$Se$_2$, (Li$_{1-x}$Fe$_x$)OHFe$_{1-y}$Se and the K-doped 3 UC FeSe/STO. Sufficient electron doping is important for this mechanism to operate fully. Although there are still different opinions we believe this mechanism is primarily due to the antiferromagnetic interaction\cite{DL}. 
\item The interaction between the FeSe electron and STO phonon. This interaction only exists in 1 UC FeSe/STO which causes extra $T_c$ enhancement in such system.
\end{enumerate}
Item (2) motivates us to propose the sandwich shown in \Fig{sand}(a) where there are two, rather than one, interfaces between STO and FeSe. Compared with FeSe/STO the phonon mediated attraction will be twice as strong. As a result $T_c$ will be more than doubled. A variant of this proposal is the bulk crystal, Fe$_2$Se$_2$SrTi$_2$O$_{5-2x}$, shown in \Fig{sand}(b) where each FeSe layer is sandwiched on both sides by TiO$_2$ planes. The structural stability (or metastability) of this and related crystals has been studied by Coh {\it et al}\cite{coh}.

In the remaining of the paper we focus on the antiferromagnetic interaction which, we believe, is the main pairing mechanism in iron-based superconductors. The first question we ask
is why Cooper pairing is so weak in bulk FeSe ($T_c\sim$ 8K). \\

In recent years many experimental and theoretical studies have concluded that the correlation strength in most iron based superconductors, with the exception of LaFePO, are intermediate to strong\cite{basov,kotliar}. In addition most of them possess fluctuating local moments\cite{gret,kotliar}. Therefore we adopt the picture that the low energy degrees of freedom include  the Fermi-surface-forming itinerant carriers and fluctuating local magnetic moments.
Because the electrons making up the local moments can interchange with those forming the Fermi surface, there is a  ``Anderson hybridization'' term between them. In such a system due to the competition between the Anderson hybridization and the direct exchange interaction between the local moments, there can be two distinct electronic phases. In the first phase the exchange interaction between the local moments dominates. As a result the local moments form a magnetic state (either long range ordered or quantum disordered) with the remaining Fermi-surface-forming itinerant carriers couple weakly to them. In heavy fermion systems this is called the ``magnetic metal'' phase. The second phase is realized when the Anderson hybridization dominates. In that case the conducting carriers have a mixed character: partly as the local-moment-forming electron and partly as the Fermi-surface-forming electron. In heavy fermion systems this is the mixed valence or the heavy Fermi liquid phase. Here the charge carriers  ``inherit'' the exchange interaction between the local moments and can form strong Cooper pairing. Apparently the optimal condition for Cooper pairing occurs near the phase transition between the two phases where the itinerant quasiparticle is poorly defined and the effective magnetic interaction is the strongest.
\\


In terms of the language used above we believe bulk FeSe is in the magnetic metal phase (although the local moments form a ``nematic quantum disordered state'' rather than a magnetic long-range ordered state\cite{wkl}.) In contrast we believe A$_{x}$Fe$_{2-y}$Se$_2$, (Li$_{1-x}$Fe$_x$)OHFe$_{1-y}$Se, the K-doped 3 UC FeSe/STO and the 1 UC FeSe/STO are in the mixed valence phase. Thus while the charge carriers do not have strong antiferromagnetic interaction in bulk FeSe, those in other four systems do. Unlike heavy fermion systems the size of the Fermi pockets in the magnetic metal and mixed valence states do not have to differ substantially. This is because the stoichiometric FeSe has two valence electrons per unit cell and the Fermi surface consists of compensated electron and hole pockets. \\
\begin{figure}
\begin{center}
\includegraphics[scale=.4,angle =0]{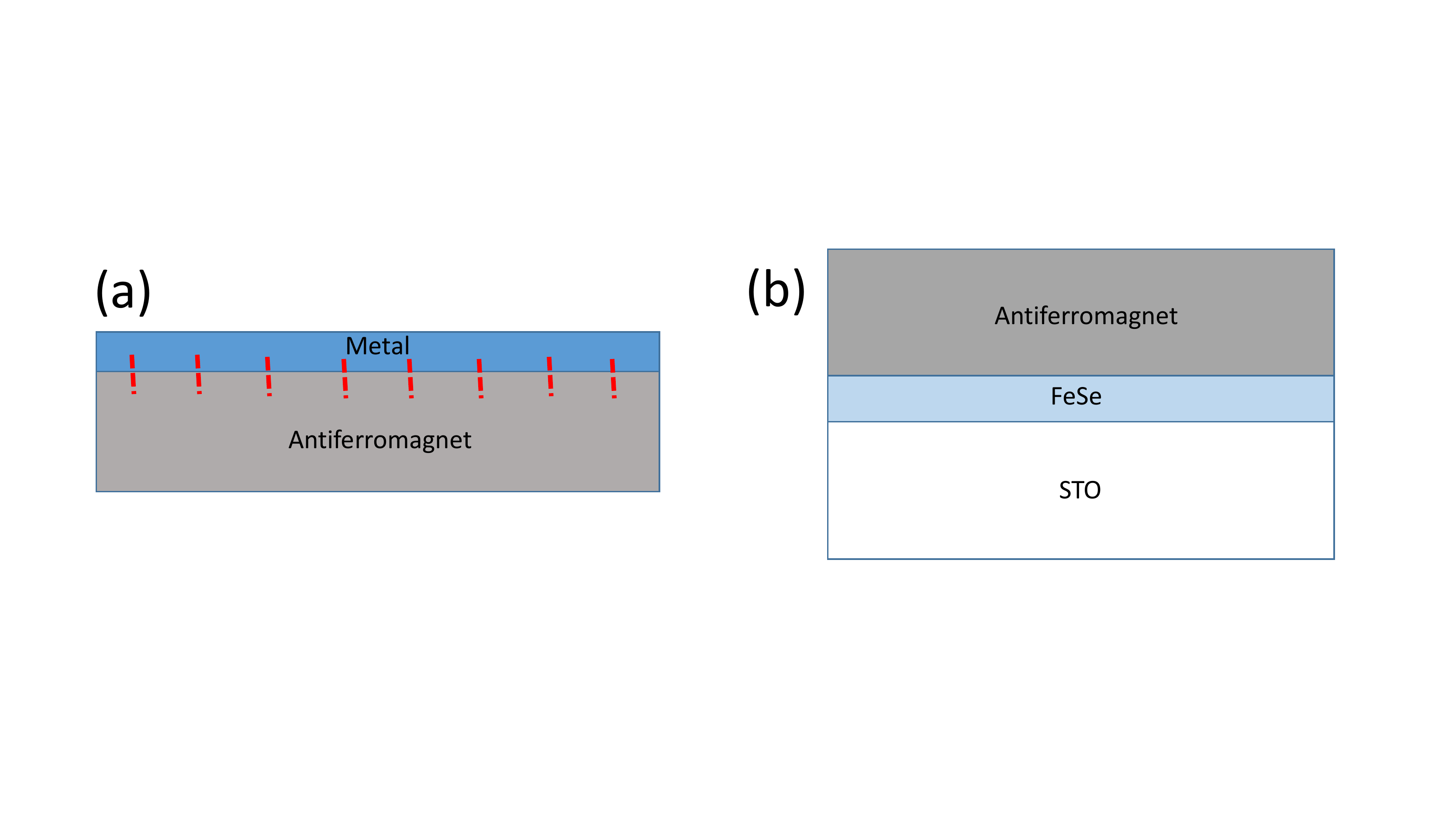}
\caption{Proposals of high $T_c$ structures. Panel (a) the interface between a metal and an antiferromagnet consisting of local moments. The red dashed lines represent single particle hybridization. Panel (b) a variant of the FeSe/STO where additional layers of local-moment antiferromagnet is grown on top of FeSe. }
\label{abc}
\end{center}
\end{figure}

The above reasoning suggests a hybrid structure where a thin metal film forms an interface with an antiferromagnet (\Fig{abc}(a)). If the Anderson hybridization is strong at the interface the carriers localized at the interface can have mixed itinerant and local moment character. This can result in strong antiferromagnetic interaction between the charge carriers and strong Cooper pairing. Thus another strategy to enhance the $T_c$ of 1 UC FeSe/STO is to deposit an antiferromagnetic insulator, e.g., La$_2$CuO$_4$, on  top of it (\Fig{abc}(b)). In this way the FeSe electrons can have the best of both worlds, nemaly, La$_2$CuO$_4$ can enhance the antiferromagnetic interaction and the STO phonon can further increase the Cooper pairing caused by the magnetic interaction.\\

{\bf Acknowledgements:}
Some of the content of this paper is already published in Ref.\cite{JJ}. It is stimulated by discussions with Zhi-Xun Shen and members of his group, Tom Deveraux and members of his group, Qikun Xue and members of his group as well as Xian-Hui Chen, Yayu Wang, Seamus Davis, Fa Wang, Steve Kivelson,  Sinisa Coh, Steve Louie and Marvin Cohen.  
DHL was supported by the U.S. Department of Energy, Office of Science, Basic Energy Sciences, Materials Sciences and Engineering Division, grant
DE-AC02-05CH11231.

\end{document}